# Price of Transparency in Strategic Machine Learning

Emrah Akyol, Cedric Langbort, and Tamer Başar
(akyol, langbort, basar1)@illinois.edu
Coordinated Science Laboratory, University of Illinois at Urbana-Champaign

*Abstract*—Based on the observation that the transparency of an algorithm comes with a cost for the algorithm designer when the users (data providers) are strategic, this paper studies the impact of strategic intent of the users on the design and performance of transparent ML algorithms. We quantitatively study the price of transparency in the context of strategic classification algorithms, by modeling the problem as a nonzero-sum game between the users and the algorithm designer. The cost of having a transparent algorithm is measured by a quantity, named here as price of transparency which is the ratio of the designer cost at the Stackelberg equilibrium, when the algorithm is transparent (which allows users to be strategic) to that of the setting where the algorithm is not transparent.

## I. INTRODUCTION

Classical Machine Learning (ML) algorithms operate under several important yet often unstated assumptions: that algorithm designer and data providers share a known common classification goal, and that they are both acting truthfully towards the goal (i.e., providers transmit truthful information, and the designer only uses it towards the stated, jointly agreed upon, goal).

However, recent technical works and investigations have shown the limits of such a paradigm, and revealed that many practical uses of ML involve, e.g., "baked-in" discrimination and/or hidden motives [1]. One approach recently proposed to limit such occurrences of invidious ML is to enforce some kind of "transparency", whereby the goal and inner-workings of the algorithm would be made public [2], [3]. Such transparency is desirable for the data providers, since it prevents, in principle, designers from acting untruthfully. However, making an algorithm public could also open the door to gaming, and adversely affect the classification outcome. In other words, there are strong incentives against transparency, even for a designer fully intent on working truthfully, and it is thus essential to quantify the possible costs for system designers, if algorithmic transparency is truly to become a widespread requirement.

The goal of this paper is to propose and compute a metric to evaluate the "price of transparency" (PoT) for specific classes of classifiers, by treating the classification task as a non-cooperative game between data providers and designer, and considering "transparency" as a widening of the common knowledge available to both players, with respect to the non-transparent version. Particularly, transparency implies that the information providers know in what context the data will be used

In this paper, we propose and analyze a metric for transparency, similar to the notion of price of anarchy (PoA) in multi-player non-cooperative games, but compares the performance attained at the (provably unique) Stackelberg equilibrium afforded by the widened transparent structure (rather than worst Nash in the PoA) to the best non-strategic attainable performance. In that sense, the models considered in this work are also similar to adversarial ML ones, with the important difference that we explicitly consider equilibria only attainable with a "transparent game structure".

Let us concretize this notion of transparency in the context of clustering (classification) algorithms in the following practical examples.

**Recommendation letter:** Consider a professor asked for a recommendation letter for a faculty job application of her graduating PhD student. The objective the search committee is to determine the candidates that would be invited for an onsite interview, based on the quality of recommendation letters. In other words, the committee infers the qualifications of the candidate based on the receiver letters, in order to make a binary decision: to invite or not invite the candidate onsite. In many cases, the professor might be biased in favor of the student since she wants her student to get the job, i.e., the objective of the professor is to render the committee's inference to be somewhat better than the candidate's actual qualifications. However, the professor is not unboundedly biased, since she does not want to compromise her reliability. Also, the bias of the professor is highly correlated with the student's actual qualifications: if she is not a good student, the professor has no incentive for bias; or alternatively, if the student is extremely successful, she does not need the professor's bias, and the professor has no incentive of risking her credibility. The committee, having received many letters, know that the letters are somewhat biased. How should the committee design their classifier (choose the candidates that will be invited for onsite interviews)?

**Dance class:** Consider a parent registering his daughter to a dance class. The information sought at the registration office, her birthday, weight, height etc, is somewhat private to the parent so he is hesitant to provide the truthful, accurate answers to such queries. The truthfulness of the answers given by the parent depends on the question asked. However, he still wants her to go to the correct dancing class (which is determined using such private information by the registration office by a transparent algorithm). Given the parent's concern about privacy conflicting with his preference that she would be assigned to correct dancing class, how should he report his daughter's information? How should the registration office assign the students to different classes?

**Number of books:** This example originates from the main motivating example of [4], but also involves a subtle but important modification: Social studies have shown that a student's success can be very well predicted from the number of books in the parents' household [5]. However, this feature cannot be used in actual school admissions simply because when it is publicly known that this is a factor in admission (if the algorithm is transparent), the parents will simply obtain several books to increase chances of their children admitted to top schools. We note however that the parents have some inference on the child's abilities: not every student can succeed in top schools; therefore, the bias (additional books bought by the parents) is not unbounded, and it may indeed be correlated with the actual qualifications of the child. In [4], this bias incurs an explicit cost to the users (the data providing agent). Here, we do not assume such a cost, but instead we assume both agents' strategies are transparent, i.e., the classifier designer knows, in a statistical sense, how much the users are misreporting.

In summary, transparency of an algorithm comes with a cost for the algorithm designer when the users (data providers) are strategic. In this paper, leveraging the recent results on strategic communication in [6], and using the well-known results in vector quantization theory, we analyze the impact of strategic intent of data provider on the design and performance of transparent ML algorithms. In other words, we analytically study the **price of transparency** in the context of strategic ML algorithms.

More formally, we study transparency in the sense of Stackelberg equilibrium of a non-zero sum information transmission game: the machinery of the algorithm (the strategy of the receiver in the information transmission game) is known to the strategic individuals (the senders of the game) who provide the training and test data. In return, the strategic intent and the functional form of the data generation (how the bias is statistically introduced into the data), are known to the ML algorithm designer. In game theory terms, this setting can be viewed as a Stackelberg signaling game between two agents: a sender (strategic data provider) and a receiver (the ML algorithm designer). The objective of the receiver is to design a classifier that would classify the samples of $X$ with minimal error (measured by a distortion metric, such as mean squared error). The objective of the sender is deceive the receiver to render it's classifier to be close to $X+\Theta$, where $\Theta$ is a bias random variable, correlated with $X$. The statistics of $X$ and $\Theta$ as well as the strategies of both agents are common knowledge (known to both agents). The data provider (sender) is restricted to pure strategies and it is the leader, and the algorithm designer (receiver) is the follower, who observes the sender's choice of pure strategies and plays accordingly to minimize its own distortion with the knowledge of the mappings of the sender. We analyze the value of this game, in conjunction with optimal strategies for both agents, to deduce the cost associated with the strategic aspect of the data provider due to the transparency of the algorithm.

## II. MODEL DESCRIPTION

In this section, we give a detailed description of our classification game and the players involved. Before discussing the strategic considerations, we give a brief technical overview of the unsupervised classification problem as well as key properties of minimum mean-squared error estimators; all results presented here are well known.

**Notation**: We let $\mathbb{R}$ and $\mathbb{R}^+$ denote the respective sets of real numbers and positive real numbers. Let $\mathbb{E}(\cdot)$ denote the expectation operator. $\mathcal{S}$ denotes the set of Borel measurable, square integrable functions $\{f : \mathbb{R} \to \mathbb{R}\}$. We use standard game theoretic notations for the related results throughout this paper (cf. [7]).

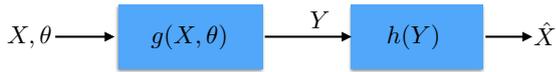

Fig. 1. Strategic Communication

### A. Clustering (Quantization, Unsupervised Learning)

Here, we review optimal clustering which is also known as the quantization or the unsupervised learning problem in different disciplines. We use the term "quantizer" to highlight the connection between machine learning and the quantization theory in which we use some well-known results. A quantizer is defined by a set of reconstruction points and a partition. The partition $\mathcal{P} = \{\mathcal{P}_i\}$ associated with a classifier is a collection of disjoint regions whose union covers $\mathbb{R}^K$. The reconstruction points $\mathcal{R} = \{r_i\}$ are typically chosen to minimize a distortion measure (such as mean squared error). The quantizer is a mapping $Q_K : \mathbb{R}^K \to \mathbb{R}^K$ that maps every vector $X \in \mathbb{R}^K$ into the reconstruction point that is associated with the cell containing $X$, i.e.

$$Q_K(X) = r_i \text{ if } X \in \mathcal{P}_i, \quad \forall i. \tag{1}$$

Throughout this paper, the optimal quantizer, that is the quantizer that minimizes MSE, is denoted by $Q_{NN}$ or the nearest neighbor quantizer.

Next, let assume that $Y$ is the strategically reported version of the source $X$. $Y$ can be viewed as the output of a channel, $p(Y|X)$ whose input is $X$. $Y$ is to be classified by a $k$-level quantizer $Q$ such that the mean squared distortion $\mathbb{E}\{||X - Q(Y)||^2\}$ is as small as possible. This $k$ level quantizer is characterized by its reconstruction points $\{y_1, \ldots, y_k\} \subset \mathbb{R}^K$ and the measurable sets $P_i = \{x \in \mathbb{R}^k : Q(x) = y_i\}, i = 1, \ldots, k$ which are partitions. It is well known (see e.g., [8]) that the following useful decomposition holds for any $Q$:

$$\mathbb{E}||X - Q(Y)||^2 = \mathbb{E}||X - \hat{X}||^2 + \mathbb{E}||\hat{X}(Y) - Q(Y)||^2 \tag{2}$$

where $\hat{X}(Y) = \mathbb{E}\{X|Y\}$. Thus, to minimize $E\{||X - Q(Y)||^2\}$, the quantizer has to minimize $\mathbb{E}\{||\hat{X}(Y) - Q(Y)||^2\}$. Therefore, the quantizer minimizing[1] $\mathbb{E}||X - Q(Y)||^2$ is obtained by first transforming $Y$ to $\hat{X} = \mathbb{E}\{X|Y\}$ and then quantizing $\hat{X}$ by a nearest neighbor quantizer $Q$. Moreover, the overall distortion can be written as the sum of two terms:

$$\mathbb{E}\{||X - Q(Y)||^2\} = D_{EST} + D_Q$$

[1]Such a minimizer exists, see e.g., [9].

where $D_{EST} = \mathbb{E}\{||X - \hat{X}||^2\}$ is the estimation distortion, and $D_Q = \mathbb{E}\{||\hat{X} - Q(\hat{X})||^2\}$ is the distortion associated with the quantization operation. Throughout this work, we choose $K = 1$ (use scalar quantizers) for the ease of exposition, although, in principle, our results are valid for any dimension, $K$. The set of $k$-level classifiers (that is, the set of scalar quantizers with $k$ partitions) is denoted here by $\mathcal{Q}^{(k)}$.

In general, the optimal quantization map is highly nonlinear and intractable. Hence, there are several common assumptions to facilitate analysis in quantization theory, one of which will be used in the sequel.

**Assumption 1.** *Quantization distortion of a random variable $X$,*

$$D_Q^{(k)}(\sigma_X^2) = \min_{Q \in \mathcal{Q}^{(k)}} \mathbb{E}\{(X - Q(X))^2\}$$

*is only a function of its variance $\sigma_X^2$ and the number of partitions ($k$) of the quantizer.*

This assumption, also known as the high resolution (high rate) quantization theory [10], holds for asymptotically high number of clusters and dithered quantizers. In practice, it is a good model for medium to high resolution for most signals and it is commonly used to analyze quantization distortion in filterbanks etc.

In this paper, we focus on scalar Gaussian variables, for which the optimal quantization rules and MSE distortions are well-studied and reported in e.g., Table 1 in [11].

### B. Strategic Communication

There exists a substantial amount of literature in information economics on communicating information in senser-receiver games: using as advertising [12], [13], education [14], disclosure of verifiable information [15], or cheap talk [16], or information disclosure[17], [18]. Here, we adopt the information disclosure model in [17] and [18] since it fits best to the transparent strategic classification setting at hand.

More formally, we consider a Stackelberg game (as opposed to Nash equilibria considered in the classical literature in Information Economics [16]), where the sender is the leader and the receiver is the follower. The game proceeds as follows: the sender plays first and announces an encoding mapping. As a leader in Stackelberg game, the sender is *committed* to its encoding mapping, i.e., the sender cannot change it after the receiver plays. The receiver, knowing this commitment, determines its own mapping that maximizes its pay-off, given the encoding map. The sender, of course, will anticipate this, and pick its map accordingly. Note that there is a natural order in this model of communication: the sender cannot change its mapping after the receiver announces the decoding strategy. Hence, a Stackelberg model is a better fit to such kind of communication systems with a natural order of communication, rather than Nash equilibria where the transmitter is not committed to its own mapping and can change it after the receiver announces the decoding mapping.

In this paper, we focus on the general communication system whose block diagram is shown in Figure 1. The source $X$ and bias $\Theta$ are mapped into $Y \in \mathbb{R}$, via a stochastic mapping $Y = g(X, \Theta)$ so that

$$\mathbb{P}(g(X, \Theta) \in \mathcal{Y}) = \int_{y' \in \mathcal{Y}} p(y'|x, \Theta) \, dy' \quad \forall \mathcal{Y} \subseteq \mathbb{R} \tag{3}$$

holds almost everywhere in $X$ and $\Theta$. Let the set of all such mappings be denoted by $\Gamma$ (which has a one-to-one correspondence to the set of all the conditional distributions that construct the transmitter output $Y$).

We take the source and bias variables jointly Gaussian, i.e., $(X, \Theta) \sim \mathcal{N}(0, R_{X\Theta})$ where, without any loss of generality, $R_{X\Theta}$ is parametrized as

$$R_{X\Theta} = \sigma_X^2 \begin{bmatrix} 1 & \rho \\ \rho & r \end{bmatrix},$$

with $r > \rho^2$. The receiver produces an estimate of the source $\hat{X}$ through a mapping $h \in \mathcal{S}$ as $\hat{X} = h(Y)$. The objective of the receiver is to minimize

$$D_R = \mathbb{E}\left\{\left(X - \hat{X}\right)^2\right\} \tag{4}$$

while that of the sender is to minimize

$$D_S = \mathbb{E}\left\{\left(X + \Theta - \hat{X}\right)^2\right\} \tag{5}$$

over the mappings $g(\cdot, \cdot) \in \Gamma, h(\cdot) \in \mathcal{S}$. We note that the sender can augment its map with any other invertible function and the receiver can do the reverse and achieve the same costs, i.e., the mappings $F(g(X, \theta))$ and $F^{-1}(h(Y))$ yield the same costs as $g(X, \theta)$ and $h(Y)$, where $F(\cdot)$ is any invertible function. This function corresponds to different permutations of labels if the message space is finite as assumed in most prior work in the Economics literature[16], [17], [18]. To account for such trivially equivalent pairs of mappings, we use the term "essentially unique", defined explicitly in the following:

**Definition 1** (Essential Uniqueness). *A mapping $g : \mathbb{R} \times \mathbb{R} \to \mathbb{R}$ is essentially unique if it is unique upto bijective transformations.*

The optimal strategies for this communication game are derived in [6] and reproduced below:

**Theorem 1** ([6]). *The essentially unique mappings at the Stackelberg equilibrium are given as $g^*(X, \theta) = X + \alpha\theta$ and $h^*(Y) = \kappa Y$, where $\alpha$ and $\kappa$ are:*

$$\alpha = \frac{A-1}{2(r+\rho)}, \quad \kappa = \frac{1+\alpha\rho}{1+\alpha^2 r + 2\alpha\rho} \tag{6}$$

*Costs at this Stackelberg solution are*

$$D_S = \sigma_X^2 \left(1 + \frac{(A-3)(r+\rho)}{A-1}\right) \tag{7}$$

$$D_R = \sigma_X^2 \left(\frac{(r-\rho^2)(A-1)}{A(2r + A\rho + \rho)}\right) \tag{8}$$

*where $A = \sqrt{1 + 4(r+\rho)}$.*

### C. Our Model

We consider a Stackelberg game between the data provider and the clustering algorithm designer. The data provider (Sender) is the leader and algorithm designer (Receiver) is the follower. The objective of the receiver is to classify the samples of $X$, given a fixed number of partitions, denoted here as $k$, with minimal MSE over the signal $X$. The objective of the sender (user) is deceive the receiver to render the classifier output close to $X + \Theta$ where $\Theta$ is a bias random variable, correlated with $X$. The statistics of $X$ and $\Theta$ as well as the strategies of both agents are common knowledge (known to both agents, due to the transparency of the algorithm).

We consider here quadratic objectives: the objective of Receiver is to minimize

$$J_R = \mathbb{E}\left\{(X - h(Y))^2\right\} \tag{9}$$

over the designed $k$-level classification map $h : \mathbb{R} \to \{r_1, \ldots, r_k\}$, while that of Sender is to minimize

$$J_S = \mathbb{E}\left\{(X + \Theta - h^*(Y))^2\right\} \tag{10}$$

over $Y = g(X, \Theta) \in \Gamma$, where $h^*$ denotes the optimal classifier mapping as a function of $g(\cdot, \cdot)$, i.e.,

$$h^*(g(X, \Theta)) = \underset{h \in \mathcal{Q}^{(k)}}{\arg\min} \, \mathbb{E}\{(X - h(g(X, \Theta)))^2\}$$

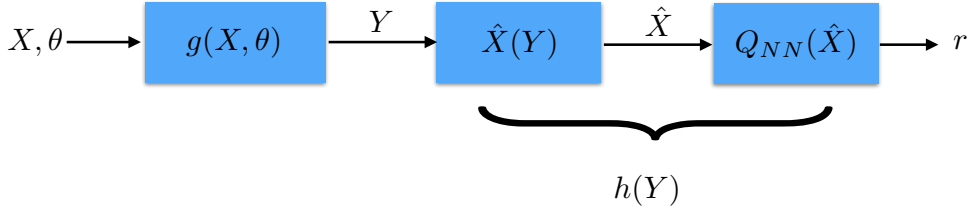

Fig. 2. The strategic classification game with the optimal classifier map.

that holds for any given $g(X,\theta)$. We let $g^*(X,\Theta)$ denote the optimal mapping for the data provider, i.e.,

$$g^*(X,\Theta) = \arg\min_{g\in\Gamma} \mathbb{E}\left\{(X+\Theta-h^*(g(X,\Theta)))^2\right\}. \quad (11)$$

Next, we define the Price of Transparency ($PoT$).

**Definition 2** (Price of Transparency). *Price of transparency ($PoT$) is the ratio of the clustering costs at the Stackelberg equilibrium (i.e., the algorithm is transparent) and the non-strategic setting (i.e., the algorithm is not transparent):*

$$PoT = \frac{J_R^*}{J_R^{**}},$$

*where*

$$J_R^* = \min_{h\in\mathcal{Q}^{(k)}} \mathbb{E}\left\{(X-h(Y))^2 \,|\, Y = g^*(X,\Theta)\right\}$$

*and*

$$J_R^{**} = \min_{h\in\mathcal{Q}^{(k)}} \mathbb{E}\left\{(X-h(Y))^2 \,|\, Y = X\right\},$$

*and $g^*(X,\Theta)$ is given in (11).*

### III. MAIN RESULTS

Before the characterization and numerical analysis of PoT, we first analyze the structure of the optimal strategies of the classifier designer and the data provider (user).

#### A. Optimal Strategies

The following theorem states the strategies achieving the Stackelberg equilibrium.

**Theorem 2.** *There exists a unique Stackelberg equilibrium with the following data providing map*

$$g^*(X,\Theta) = X + \alpha\Theta, \quad (12)$$

*where*

$$\alpha = \frac{-1+\sqrt{1+4(r+\rho)}}{2(r+\rho)},$$

*and the following classifier*

$$h(\cdot) = Q_{NN}(\cdot) \circ \hat{X}(\cdot)$$

*where $\hat{X}(Y) = \mathbb{E}\{X|Y = g^*(X,\Theta)\}$ and $Q_{NN}(\cdot)$ is the optimal (nearest neighbor) classifier.*

*Proof:* The optimal strategy for the classifier follows from (2), i.e., first estimate $X$ from $Y$ via MMSE estimation, denoted here as $\hat{X}(Y)$, and then quantize $\hat{X}(Y)$ with the optimal nearest neighbor classifier $Q(\hat{X})$, designed for $\hat{X}$. The data provider's cost, in conjunction with this classifier, $J_S$ can be expressed as:

$$J_S = \mathbb{E}\{(X+\Theta-\hat{X}(Y))^2\} + \mathbb{E}\{(\hat{X}(Y)-Q(\hat{X}(Y)))^2\} + 2\mathbb{E}\{(\hat{X}(Y)-Q(\hat{X}(Y)))(X+\Theta-\hat{X}(Y))\} \quad (13)$$

The first term above, $D_S = \mathbb{E}\{(X+\Theta-\hat{X}(Y))^2\}$, is the sender cost of the communication game outlined in Section II-B. Theorem 1 implies that $g^*(X,\Theta)$ in (12) is the unique minimizer of $D_S$.

The last term in (13) is the sum of two terms $\mathbb{E}\{(\hat{X}(Y)-Q(\hat{X}(Y)))(X-\hat{X}(Y))\}$ and $\mathbb{E}\{\Theta(\hat{X}(Y)-Q(\hat{X}(Y)))\}$. The first one vanishes since estimation error $X - \hat{X}(Y)$ is orthogonal to any deterministic function of the observation $Y$. The second one can be expressed as in (14). where $f_\Theta(\cdot)$ is the density of $\Theta$. Noting the fact that optimal quantizer (classifier) is an unbiased estimator, i.e., $\mathbb{E}\{Q(\hat{X}(Y))|\Theta\} = \mathbb{E}\{\hat{X}(Y)|\Theta\}$ almost surely in $\Theta$, we conclude that this term also vanishes.

The second term $\mathbb{E}\{(\hat{X}(Y) - Q(\hat{X}(Y)))^2\}$ is the quantization error of the estimate $\hat{X}(Y)$. Note that $g^*(X,\Theta)$ in the theorem statement minimizes $D_S(\sigma_{\hat{X}}^2)$ subject to fixed $D_R$. Fixing $D_R(\sigma_{\hat{X}}^2)$ implies that $\sigma_{\hat{X}}^2$ is fixed since,

$$D_R(\sigma_X^2) = \sigma_X^2 - \sigma_{\hat{X}}^2$$

$$\mathbb{E}\{(\hat{X}(Y)-Q(\hat{X}(Y)))\,\Theta\} = \int \mathbb{E}\{(\hat{X}(Y)-Q(\hat{X}(Y)))|\Theta = \theta\}\theta f_\Theta(\theta)\mathrm{d}\theta \qquad (14)$$

by the basic orthogonality principle of estimation. Since $\mathbb{E}\{(\hat{X}(Y)-Q(\hat{X}(Y)))^2\} = D_Q^{(k)}(\sigma_{\hat{X}}^2)$ is only a function of the estimate variance $\sigma_{\hat{X}}^2$ (by Assumption 1) which is fixed, we conclude that $g^*(X,\Theta)$ minimizes $J_S$, which implies the optimality of $g^*(X,\Theta)$ when used in conjunction with the optimal $h(\cdot)$.

### B. Characterization of PoT

In the following, we present a characterization of PoT for our setting.

**Theorem 3.** *The price of transparency is*
$$PoT = \frac{D_R(\sigma_X^2) + D_Q^{(k)}(\sigma_{\hat{X}}^2)}{D_Q^{(k)}(\sigma_X^2)},$$
*where $\sigma_X^2$ is the source variance and $\sigma_{\hat{X}}^2$ is the variance of the receiver's estimate in the strategic communication game.*

*Proof:* The value of PoT can simply be found by plugging the relevant expressions.

### IV. CONCLUSIONS AND DISCUSSIONS

This paper constitutes a first attempt to quantitatively analyze the cost of transparency in ML algorithms, specializing to classification/clustering type of ML problems. We have conducted a game theoretical analysis of strategic classification/clustering for a jointly Gaussian source-bias model. Building on the recent prior work on strategic communication, we have shown how the price of transparency changes with the joint statistics of bias and source, and also the number of classification regions. Several practical problems are, including an investigation of optimal strategies if the source statistics is not available apriori but estimated from the strategically generated training samples? The quantization theory provides some starting points for this problem [19], [20], which will be pursued as a part of our future work.